# Learning from Nature to Improve the Heat Generation of Iron-Oxide Nanoparticles for Magnetic Hyperthermia Applications

Carlos Martinez-Boubeta[1,*], Konstantinos Simeonidis[2], Antonios Makridis[3], Makis Angelakeris[3], Oscar Iglesias[4], Pablo Guardia[5], Andreu Cabot[1,5], Lluis Yedra[1], Sonia Estradé[1,6], Francesca Peiró[1], Zineb Saghi[7], Paul A. Midgley[7], Iván Conde-Leborán[8], David Serantes[8] & Daniel Baldomir[8]

[1]Departament d'Electrònica and IN2UB, Universitat de Barcelona, Martí i Franquès 1, 08028 Barcelona, Spain.

[2]Department of Mechanical Engineering, University of Thessaly, 38334 Volos, Greece

[3]Department of Physics, Aristotle University of Thessaloniki, 54124 Thessaloniki, Greece

[4]Departament de Física Fonamental and Institute IN2UB, Universitat de Barcelona, Av. Diagonal 647, 08028 Barcelona, Spain

[5]IREC, Jardins de les Dones de Negre 1, 08930 Sant Adrià del Besòs, Spain.

[6]TEM-MAT, CCiT-Universitat de Barcelona, Solé i Sabaris 1, 08028 Barcelona, Spain.

[7]Department of Materials Science and Metallurgy, University of Cambridge, Pembroke Street, Cambridge CB2 3QZ, United Kingdom.

[8]Instituto de Investigacións Tecnolóxicas, and Departamento de Física Aplicada, Universidade de Santiago de Compostela, 15782 Santiago de Compostela, Spain.

*E-mail: cmartinezboubeta@ub.edu; daniel.baldomir@usc.es

**The performance of magnetic nanoparticles is intimately entwined with their structure, mean size and magnetic anisotropy. Besides, ensembles offer a unique way of engineering the magnetic response by modifying the strength of the dipolar interactions between particles. Here we report on an experimental and theoretical analysis of magnetic hyperthermia, a rapidly developing technique in medical research and oncology. Experimentally, we demonstrate that single-domain cubic iron oxide particles resembling bacterial magnetosomes have superior magnetic heating efficiency compared to spherical particles of similar sizes. Monte Carlo simulations at the atomic level corroborate the larger anisotropy of the cubic particles in comparison with the spherical ones, thus evidencing the beneficial role of surface anisotropy in the improved heating power. Moreover we establish a quantitative link between the particle assembling, the interactions and the heating properties. This knowledge opens new perspectives for improved hyperthermia, an alternative to conventional cancer therapies.**

**SUBJECT AREAS**: NANOSCALE MATERIALS. MAGNETIC PROPERTIES AND MATERIALS. THEORY AND COMPUTATION. NANOTECHNOLOGY IN CANCER

Consider an assembly of single-domain particles with uniaxial anisotropy, coupled to each other by the magnetic dipole-dipole interaction. Such interactions can be tuned by adjusting the size, the magnetization, and the volume fraction of the particles[1]. Even in the superparamagnetic regime, the collective magnetic behavior will differ from that of isolated particles[2]. A potentially interesting area where this effect may find applications is biomedicine. An example of the latter is the optimized design for multicore particles achieving enhanced transverse relaxivities for magnetic resonance imaging[3,4]. Yet magnetic nanoparticle suspensions have gained an important role in cancer treatment with AC hyperthermia. In synergy with chemotherapy or radiotherapy, selective targeting and localized heating of tumor cells can be tuned, leading to modalities with shorter time regimes even with a lower dosage[5]. Indeed, the efficiency of this type of radio-frequency thermotherapy has been demonstrated on several types of cancers including brain tumor, prostate cancer, and invasive breast carcinoma[6]. Although encouraging results on palliative care indicate that even non-optimized particles with the appropriate size distribution can deliver adequate heating power if present in sufficiently high concentrations[7], concerns have been raised regarding the toxicity for cancer-directed therapy[6]. In order to minimize the potential side effects arising during the clinical treatments, the quantity of nanoparticles administered needs to be as small as possible but still retaining the desired effect (see Supplementary Information). For this purpose, to reach the therapeutic temperature with minimal particle concentration in tissue, the magnetic nanoparticles should exhibit high inductive specific absorption rate (SAR).

This quantity depends on the nanoparticles' properties, such as mean size, saturation magnetization ($M_S$) and magnetic anisotropy (K), but also on the alternating magnetic field amplitude ($H_{max}$) and frequency (*f*). In previous work[8], heating has been predicted for superparamagnetic nanoparticles within a model in which SAR primarily depends on magnetic spin relaxation processes. It was shown that the crossover between Néel and Brown regimes of relaxation depends on the anisotropy constant and particle volume[9], thus defining, for each frequency, a narrow range of K and size values for optimal SAR. Furthermore, note that the aforementioned model is suitable to calculate SAR only for very small nanoparticles in the diluted regime at low magnetic fields[10], whereas for clinical applications the relative influence among particles cannot be neglected. For instance, results for dense systems have shown that dipolar interactions not only affect the susceptibility[11,12], but also the blocking temperature transition[13], and the motion of particles in solution[14]. Hence, the selection of the most advantageous materials for clinical hyperthermia treatment is still a matter of debate[6].

Previously, the tuning of the magnetocrystalline anisotropy and its influence on magnetic heating efficiency has been issued by Lee *et al.*[15] in exchange-coupled nanoparticles. Note that K can also be controlled by changing the nanoparticle shape. Yet, another interesting option to improve the heat generation from magnetic nanoparticles could be to explore single domain particles displaying hysteresis losses[16], especially if one considers the SAR reduction usually found in the small superparamagnetic particles after immobilization (for instance, into cells)[17]. Magnetization reversal calculations within the Stoner-Wohlfarth model for independent, randomly oriented uniaxial single-domain particles, lead to losses scaling approximately with $\mu_0 M_S H_C$, being $H_C$ their coercive field. In this regard, metallic iron-based nanoparticles (instead of iron-oxides) are the best candidates as they virtually present the highest saturation magnetization. For instance, iron nanocubes[18] (with effective anisotropy constant $K_{eff} = 9.1 \times 10^4$ J/m$^3$, and $M_S = 1.7 \times 10^6$ A/m) display the highest specific losses reported in the literature so far, provided that the field amplitude employed is sufficiently high to ensure the remagnetization of the particle (SAR about 3000 W/g at $\mu_0H_{max}$ = 73 mT and $f$ = 274 kHz). Nevertheless, soft magnetic nanoparticles seem to be preferable for the purpose of hyperthermia within the range of magnetic fields used in clinical trials[19]. On the other hand, the low chemical stability at physiological conditions and adequate biocompatibility are drawbacks of metallic particles, thus forcing the combinatory use of complex core-shell architectures[20,21]. Consequently, up to date, the majority of investigations focused on magnetic iron oxides $Fe_3O_4$ (magnetite) and $\gamma$-$Fe_2O_3$ (maghemite), which have been proved to be well tolerated by the human body. Therefore, it is highly desirable to determine hyperthermia efficiency of iron oxide particles with mean size ranging above the superparamagnetic limit ($\geq$ 15 nm) and below the optimal size ($\leq$ 50 nm) for internalization into mammalian cells[22], especially if one has in mind the case of intracellular hyperthermia[17].

Inspired by the excellent heating properties of bacterial magnetosomes having a mean diameter of the magnetite crystals of about 30 nm[23,24], here we demonstrate that cubic nanoparticles possess superior heat induction power compared to spherical iron oxide particles of similar size. This higher hyperthermia performance is mainly related to the higher surface induced magnetic anisotropy and the tendency towards aggregation into chains facilitated by the cubic shape. Although previous experimental studies of the magnetic properties of cubic particles have been done[25,26], showing that cuboids could be good materials for the biomedical imaging applications[27,28], little has been done to compare the effects of spherical versus cubic particles in hyperthermia[29]. Some of us have previously reported dependence of the SAR on the mean particle size[30,31]. Based in our know-how in the size and shape control of iron oxides, here we provide a fundamental aspect of tuning the heat dissipation efficiency by adjusting the morphological profile of magnetic particles. For our proof-of-concept, we first synthesized, structurally and magnetically examined 20 nm cubic particles as prototype hyperthermia carriers in contrast to standard spherical ones. We have also compared the 20 nm cubic particles to cuboids with an increased size of 40 nm in order to account for thermal contributions to the energy barrier for magnetization reversal. Theses sizes were deliberately selected so that the particles safely reside within single magnetic domain range[32,33]. On the one hand, by performing systematic ac magnetic measurements in aqueous solution, we demonstrate that, the SAR is strongly related not only to the shape, the volume and the concentration of particles, but it might as well be influenced by their aggregation into chains. On the other hand, the analysis of the numerical calculated hysteresis loops allows for the correlation of their magneto-structural properties, the influence of anisotropies, and dipolar interactions, with hyperthermia measurements.

## Results

Figure 1 shows the transmission electron microscopy (TEM) view of representative aggregates of cube-shaped iron oxide particles (see Supplementary Fig. S1 for a TEM image of the spherical particles). The difference in contrast within the image is because the different crystallographic orientations of individual particles with respect to the electron beam. In this regard, there is no apparent contrast variation within each nanoparticle, thus pointing to the fact that particles are completely oxidized during synthesis[34,35]. A high-resolution (HRTEM) image clearly attests to the monocrystalline structure with a lattice fringe of 0.24 nm characteristic of (222) planes of the spinel. As can be seen in Fig. 1b (inset), the cubes organize themselves into chains with sharing flat faces of the {100} type.

We used electron tomography to generate a 3D illustration of the self-assembly of nanoparticles (see Fig. 2), similar to that described in biogenic magnetosomes. The faceted cubic particles are reconstructed separately in Fig. 2c (see Supplementary Movie S1 and Fig. S2 online for details). It should be noted here, that van der Waals attractions are expected to be weak in our nanoparticle solutions due to the thick surfactant layers (the decanoic acid $CH_3(CH2)_8COOH$ has a carbon chain length of 1.4 nm). The thermogravimetric analysis (TGA) was employed to verify the coating structure of the nanoparticles. The experimental values of weight losses (see Supplementay Fig. S4) are in reasonable agreement with the theoretical estimation assuming that the surfactant forms a close-packed monolayer on the nanoparticles[36]. Thus, the formation of such chain-like aggregates is favored because the energy of the magnetic dipole-dipole interaction is presumably larger than the thermal energy, and the van de Waals or electrostatic interactions within the solution[37].

A simple way of visualizing the consequences of these dipole interactions is to look at the magnetic response. Field and temperature-dependent magnetic measurements were performed on the dried crystalline powder using a superconducting quantum interference device (SQUID) magnetometer. Noteworthy, the remanent magnetization ($M_R$) values are well below the $0.5M_S$ (see Fig. 3, and Supplementary Fig. S5 for hysteresis loops of the spherical particles) expected for magnetically independent uniaxial Stoner-Wolhfarth nanoparticles, thus signifying non-negligible dipolar interactions between the particles. The 40 nm cubic particles are ferrimagnetic at room temperature with a saturation magnetization value up to 89 $Am^2/Kg$, and coercive field of 5.5 mT. The smaller the particles, the smaller the magnetization of the sample is, probably because of the appearance of cation vacancies and surface spin canting on decreasing the particle size[38,39,40]. Hysteresis loops also indicate a reduction in coercivity as the mean particle size is decreased, and an increase of both $M_S$ and $H_C$ as the temperature is reduced from 300 K (Fig. 3b). Both observations are consistent with a lowering of the energy barrier for magnetization reversal that leads to faster relaxation by thermal fluctuations[41]. Importantly, data suggest that the 20 nm particles at room temperature are in the transition regime between superparamagnetism and a blocked state. The effect of temperature upon the magnetic anisotropy is a topic for over 50 years ago[42]. We have evaluated the effective anisotropy constant from the law of approach to magnetic saturation (see the Supplementary Information for details)[43]. The parameters of these particles are summarized in Table S1. Remarkably, the $K_{eff}$ can be modulated by varying the size and shape of the nanocrystals, with values above those for bulk $Fe_3O_4$ (about 11 kJ/m$^3$) or γ-$Fe_2O_3$ (13 kJ/m$^3$) at room temperature. We deem cubic particles exhibit higher anisotropy energy values than that for spheres due to shape contribution. Note that the sphere has the smallest surface area among all surfaces enclosing a given volume. It is therefore not surprising that we find an increased anisotropy in the case of cubic particles, compared to the spherical ones. The problem, however, might be more complicated because of the possible smaller crystallinity of the spherical particles compared to cubes[44,45]. It is known that surface anisotropy is linked to well-defined lattice planes[46], being the spherical entities formed by different nanofacets while the cubic particles have fairly flat {100} planes.

The temperature dependence of zero-field-cooled and field cooled (ZFC-FC) magnetization curves establishes further differences between the nanomagnetic features of the isolated particles and the collective system due to the variations in the magnetostatic interactions[47]. For example, FC curves (see Fig. S6) reach a plateau in contrast to the random non-interacting particle system, which points out to a strong interaction between particles. The details and the mechanisms behind the power loss of particles in such a collective intricate behavior are not sufficiently understood yet[48,49]. Indeed we find significant changes in the hyperthermia response depending on size, the shape of nanoparticles and concentration, which do not follow the linear response theory[8]. Figure 4 presents the influence of particle concentration on the heat production for 20 nm and 40 nm iron oxide nanocubes in aqueous dimethyl sulfoxide (DMSO) solutions. Obviously, the available AC field-intensities probe only losses from minor hysteresis loops for the 40 nm sample (see inset in Fig. 3a), thus explaining the smaller SAR values compared to the 20 nm case. We observe also that SAR decreases with increasing concentration in both cases and that, as expected, it increases in a non-linear way with the maximum applied field for all the studied samples. Next, the heating efficiency of 20 nm iron oxide nanocubes was compared to spherical nanoparticles, considering that both systems are of the same size and crystal structure differing only in their shape. Figure 5a shows that the SAR measured under the same experimental conditions is about 20% superior for the nanocubes, despite similar room temperature soft magnetic features are also observed for the spherical nanoparticles (see Supplementary Fig. S5).

To provide insights into the experimental parameters that dictate the pronounced dependence of SAR on concentration and particle shape, we performed Monte Carlo simulations (see Fig. 5b). Our analysis is based on the macrospin approximation that considers each particle as a classical Heisenberg spin with an effective anisotropy, which can differ from the bulk first-order anisotropy constant since it includes surface anisotropy contributions[40,50,51]. Some representative simulated hysteresis loops are shown also in the Supplementary Information (see Fig. S9). Calculations endorse the statements of the previous paragraphs: a widely negative influence of dipole-dipole interactions on the heating power for an increasing particle concentration, and the greater hysteresis area for the nanocubes. In view of these observations, it is unambiguous that cubic particles exhibit higher anisotropy energy values than that for spheres. Although pioneering reports have revealed the imperative contribution of surface anisotropy to the magnetic properties of fine cubic particles, its role is still an issue of controversy. In order to gain some insight on the peculiarities of the experimental magnetic behavior presented in the previous section, we have conducted a MC simulation study of individual spherical and cubic nanoparticles in which the magnetic ions forming the particle are considered at the atomistic level of detail. We observe that the area of the hysteresis loop of the cubic particle is bigger than that of the spherical one (see Fig. 6), implying also a higher SAR as observed experimentally. Notice that the difference in areas stems from qualitative loop shape differences around the coercive and closure field points that can be traced back to the different reversal processes of the surface spins (see Supplementary Fig. S11 and S12).

## Discussion

A nice survey of literature, and comparison of experimental and calculated SAR values for several nanoparticles types as a function of the magnetic field frequency (range of 300 kHz–1.1 MHz and with an amplitude up to 31 mT) can be seen in Ref. 10 and 17. While commonly available iron oxide ferrofluids show SAR about 100 W/g[30], in a few special cases experimental values in excess of 500 W/g[52], and up to about 1 kW/g were found for magnetosomes[23,24]. Attention should be paid to the fact that the values are usually reported in Watts per iron gram, accordingly a factor about 1.5 will scale up our results for magnetite, thus concluding that our values are among the highest recorded for iron oxide nanocrystals. Naively, magnetic losses scale approximately with $M_S H_C \propto K$ given that $H_C \approx 0.5 H_A$ within the Stoner-Wohlfarth model, being $H_A = 2 K/M_S$ the characteristic anisotropy field of the particles (see Supplementary

Information). Thus, the increased anisotropy in the case of cubic particles explains the higher SAR compared to the spherical ones.

The results reported in the present study also reveal that, in addition to the surface magnetic anisotropy energy, particle concentration plays a crucial role in tailoring the heating efficiency independently of the particle geometry, as demonstrated in Fig. 4. Reasonably, an increase in concentration corresponds to a decrease in the mean interparticle separation and gives rise to a notable increase in the dipolar interparticle interactions. The role that dipolar interactions might have in SAR is not completely understood at present[53], and recent experimental studies have reported either an increase or decrease of SAR with interactions[12,16,54,55,56,57]. Overall, results suggest a widely negative influence of dipole-dipole interactions on the heating power of nanoparticles, still a clear correlation between the magnetic anisotropy of the particles and their hysteresis loop is observed and may lead to an increase in magnetic hyperthermia efficiency[58].

Additionally, the different geometrical arrangement of nanoparticles in suspensions may play some role in explaining the increase of SAR for nanocubes compared to spherical particles. For instance, our observation of nanocubes chain formation even in the absence of a magnetic field (Fig. 2) suggests the existence of strongly anisotropic dipolar forces mediating nanoparticle attachment[37]. We are not aiming to relate chain formation to concentration increase, but we argue that, since nanocubes are more prone to chaining, they would display higher SAR values compared to the spherical particles. In this regard, a rigorous computation of the magnetostatic energy of chains of identical, uniformly magnetized particles of arbitrary shape can be found in Phatak *et al.*[59]; it is shown that the face-to-face configuration in nanocubes allows for significantly more favorable chain formation than the case for spherical entities. Obviously, a transient long nanoparticle chain could also form by the application of an alternating magnetic field[14], though the self-assembly of colloidal crystals into ordered superstructures depends critically on the shape (and size) of the nanocrystal building blocks[37]. Note that, interestingly, straight chains made of cubes will have an axial magnetization state as the lowest in energy, while it is unlikely that a chain of 20 nm iron-oxide nanospheres will be observed experimentally[60]. It is worth mentioning the hydrodynamic diameters measured by dynamic light scattering (see supplemental Fig. S3). Results suggest that 20 nm cuboids self-aggregate in aqueous solution, and the average number of particles per chain is roughly 10, confirming the formation of agglomerates in these samples while the nanospheres stay dispersed.

Thus, we speculated that the capacity of cuboids to self-assemble spontaneously can be used to tailor the heating capabilities. To further cross check this view, we have computed the evolution of the hysteresis as a function of the number of particles N within a chain. We will focus now on the influence of interparticle dipole interactions in a chain arrangement as displayed in Fig. 7. Our simulations predict that the area of hysteresis loop increases (and therefore of the SAR) with the length of the chain. Further, the thermal stability gained by creating arrays is an advantage when exploiting hysteresis losses (inset Fig. 7). These observations indicate a promising way to increase the hyperthermia performance by assembling cubic particle in elongated chains. Independently of the formation of chains and its positive effects on SAR, a concentration increase may lead to the coalescence of chains leading to cluster formation and a decrease in SAR. However, we stress that even in this case, the higher SAR of cubes can be account by their increased surface anisotropy with respect to spheres. This finding is in remarkably good agreement with the results observed for magnetosomes by Alphandéry *et al.*[61] Disorientation of the assembly would lead to a considerable decrease in the hysteresis loop area and to very different heating properties. Example of the later is the decrease on the heating efficiency of separate magnetosomes compared with those of magnetosomes arranged in chains[23,24].

To conclude, our data and analysis indicate that ferrimagnetic nanocubes with an edge length about 20 nm exhibit superior magnetic heating efficiency compared to spherical particles of similar sizes. The oriented attachment of magnetic nanoparticles biomimicking magnetostatic bacteria, and the beneficial role of surface anisotropy, are recognized as important mechanisms for the development of magnetic hyperthermia for cancer treatment. We foresee such quantification of nanoparticles interaction and understandings of the magnetization reversal are also important for the design of magnetic nanomaterials for other biomedical applications.

## Methods

**Synthesis.** Iron oxide nanocubes were prepared by heating a solution of iron(III) acetylacetonate (Fe(acac)$_3$), decanoic acid and dibenzylether. This method takes the advantage of the discernible separation of nucleation and growth stages caused by the intermediate formation of iron(III) decanoate complex as discussed in detail in a previous publication[62]. Briefly, size can be tuned over a wide range (15 nm to 180 nm) by choosing the suitable heating rate. Namely, for the preparation of 20 nm nanocubes (edge dimension), 0.353 g (1 mmol) of Fe(acac)$_3$ was mixed with 0.688 g (4 mmol) of decanoic acid in 25 mL of dibenzyl ether. After a short vacuum step at 60 ºC (30 minutes), the solution temperature was first raised up to 200 ºC with a constant rate of 2.6 ºC/min, and kept at this temperature for 2 h under an argon flow and vigorous stirring. In a second step, the solution was heated to reflux temperature with a heating rate of 6 ºC/min. After 1 h the solution was cooled down and acetone was added. Nanoparticles were then collected by centrifugation at 8000 rpm and redispersed in chloroform. This procedure was repeated at least two times in order to get rid of the excess of surfactant. Nanocubes of 40 nm were synthesized by decreasing the heating rate, during the second step, down to 1.7 ºC/min. In the synthetic procedure for the production of 20 nm (diameter) spherical particles, 3 mmol Fe(CO)$_5$ were added at 100 ºC in 10 mL dioctylether in the presence of 12 mmol oleic acid. The mixture was left to reflux for 3 h at 290 ºC and then cooled to room temperature. Ethanol was added to yield a black precipitate, which was then separated by centrifugation. The supernatant was discarded and the particles were redispersed in hexane. Given that the resulting magnetic nanoparticles were hydrophobic, the powders were further dissolved in a mixture of dimethyl sulfoxide (DMSO) and water (1:1). DMSO is a naturally derived, inexpensive, non-toxic solvent and pharmaceutical agent that has been demonstrated to be a well-tolerated excitatory modulator in the management of cancer pain.[63] Furthermore, DMSO has chemical properties which facilitate its absorption into and distribution throughout biological systems by all routes of administration, thus biocompatibility in future applications is guaranteed.

**Characterization.** Iron oxide particles were characterized by transmission electron microscopy (TEM) using a JEOL JEM-2100 (200 keV) for high resolution (HR)TEM, and by a field-emission gun scanning transmission electron microscope (FEG)TEM FEI Tecnai F20-G2 (200kV) equipped with a high-angle annular dark-field detector STEM-HAADF for three-dimensional (3D) electron tomography. The samples were prepared by dropping on solution of nanoparticles onto a carbon coated copper grid. The 3D structure of nanoparticles, as well as that of the assembled chain, is reconstructed from a tilt series (range of -64º ≤ α ≤ 64º at an increment of 4 degrees) of 2D projections, using the simultaneous iterative reconstruction technique (SIRT). In order to determine the apparent hydrodynamic diameter, dynamic light scattering (DLS) analysis was carried using Malvern Instruments Hydro 2000MU accessory. Thermogravimetric analysis (TGA) was carried out with a TGA-SDTA 851e/SF/1100 Mettler Toledo device up to 700 ºC, by heating the sample under an argon/nitrogen flow at a heating rate of 10 ºC/min. The residual weight accounts for the mass of iron-oxide nanoparticles in the ferrofluid. Quasi-static magnetic characterization was carried out in a superconducting quantum interference device (SQUID) Quantum Design MPMS XL-7T magnetometer. M(H) hysteresis loops were measured at different temperature (5K and 300K) by applying field up to 5 Tesla. Magnetization zero field cooled (ZFC) measurements were performed upon warming with an applied magnetic field of 5 mT after cooling the samples in zero applied field. The field cooled (FC) curves were obtained by measuring at stepwise-decreasing temperatures in the same small applied field. AC magnetic hyperthermia experiments were performed using a 23 mm diameter three-turn induction coil powered by a 4.5 kW AC field generator. While frequency was kept constant at 765 kHz, the amplitude of the applied magnetic field was tuned from 15 up to 30 mT. The amplitude of the alternating magnetic field was estimated using a pick-up coil connected to an oscilloscope. Temperature was monitored by using an OpSens PicoM GaAs-based fiber optic probe immersed in a test tube containing 2 mL of solution. Three different iron-oxide concentrations (0.5, 1.0 and 2.0 mg/ml) were exposed to the alternating magnetic field for 900 seconds. Specific absorption rate (SAR) values were estimated by subtracting the solvent background signal and the heat losses to the environment. Further details on the hyperthermia capabilities of these particles under safe clinical conditions ($H_{max}f < 5 \times 10^8$ A/ms), and intracellular uptake trials, can be found in a recent publication[31].

**Computational details.** The simulations were performed both at the atomic level and under the so-called macrospin approximation, in order to investigate the single-particle properties in relation to shape and anisotropy, and the role of magnetic dipolar interactions, respectively. First, we used the Monte Carlo (MC) method with the standard Metropolis algorithm. The physical model employed for our numerical simulations considers a perfectly monodisperse system of single-domain magnetic particles with an effective uniaxial magnetic anisotropy. Though, it should be pointed out that the SAR may be also influenced by the particle size distribution (as recently considered by Hergt *et al.*[64] and Carrey *et al.*[10]), which, however, is beyond the scope of the present paper. The spatial distribution of the particles resembles a frozen ferrofluid without aggregations, the positions of the particles are kept fixed and the easy axes are chosen randomly. The energy model is the same as in Ref. 48 so that the energy of each particle in this ideal scenario is regarded to have three main sources: anisotropy, Zeeman and dipolar interaction. In our simulations, we reproduce M(H) curves at different sample concentrations and for different values of maximum applied field $H_{max}$. To fit the simulations with the experimental procedure, we applied magnetic fields of the same amplitude as were applied experimentally. The model shows that SAR arrives at the summit for large field amplitudes, for values about the characteristic anisotropy field of the particles $H_A = 2 K/M_S$. Next, we have considered an atomistic model with Heisenberg classical spins placed at the magnetic ion lattice sites of the real maghemite structure (see Supplementary Information for details about the model and simulations techniques), and we have computed the low temperature hysteresis loops of two individual particles with spherical and cubic shapes having the same size as the ones experimentally studied. Finally, in order to investigate the degree of magnetostatic coupling between nanoparticles, we studied a series of chains with N particles. The anisotropy axis of the particles was arbitrary taken within the chain axis, equally distributed apart from the median position in order to account for deviations in the positions of the crystals from the chain axis, thus resembling the results obtained by electron holography on magnetosomes[65].

**Acknowledgments**

This work was co-financed by the Spanish MAT2009-08667, MAT2009-08165 and SAF2011-25707 projects. C. M. Boubeta and A. Cabot were supported by the Spanish Government under the 'Ramón y Cajal' Fellowship program. I. Conde-Leborán was supported by the FPI Spanish program. K. Simeonidis acknowledges "Education and Lifelong Learning" Operational Program funded by EU-European Social Fund (ESF) and GSRT. Z. Saghi and P. A. Midgley acknowledge financial support from the European Union within the Framework 6 program under a contract for an Integrated Infrastructure Initiative, Reference 026019 ESTEEM. P. A. Midgley also acknowledges financial support from the European Research Council, Reference 291522-3DIMAGE. O. Iglesias acknowledges funding from the European Union FEDER funds ("Una manera de hacer Europa") and Catalan DIUE through Project 2009SGR856. He also acknowledges CESCA and CEPBA under coordination of C4 for supercomputer facilities. We would like to thank A. Matsakidou and Prof. V. Kioseoglou from Laboratory of Food Chemistry and Technology (AUTh) for the DLS measurements, and also the Centro de Supercomputación de Galicia (CESGA) for the computational facilities.

**Contributions**

C.M.B. suggested the study. C.M.B., M.A. and D.B. led the project. P.G. and K.S. fabricated the samples and performed the TGA and SQUID measurements. L.Y., S.E., F.P., Z.S. and P.A.M. led the collection and the analysis of the electron microscopy images and 3D tomography. I.C.L., D.S., and O.I. performed and interpreted the Monte Carlo simulations. K.S. and A.M. were responsible for the hyperthermia measurements. C.M.B., K.S., M.A., O.I., D.S. wrote the manuscript with substantive feedback from A.C., S.E. and D.B.

**Competing financial interests**

The authors declare no competing financial interests.

**Figure Legends**

**Figure 1. TEM images**. Iron-oxide nanocubes (a) with averaged edge length of 20 ± 4 nm; inset reveals 2D self-assembly arrangements. (b) Corresponding TEM micrograph of 40 nm nanocubes. As can be seen in the larger area view, the particles organize themselves in different chain-like configurations. (c) High-resolution observation of crystal structure revealing (222) fringes of the inverse spinel iron oxide.

**Figure 2. Tomography.** 3D reconstruction of cuboctahedral shape particles, from images obtained at different tilt angles relative to the electron beam, after 40 iterations (see Methods). (a) Nanocube cluster. Neighboring nanocubes have {100} surfaces face to face separated by a distance above 2 nm due to hydrocarbon ligands. (b) Single nanocube in its original context. (c) Illustration of small deviations from perfect cubic symmetry.

**Figure 3. SQUID.** (a) Magnetic hysteresis loops recorded at 300 K for 20 nm and 40 nm square nanoparticles. Inset shows the low field region of the hysteresis loop. (b) Magnetic hysteresis loops recorded at 5 K.

**Figure 4. Hyperthermia.** Specific Absorption Rate (SAR) for (a) 40 nm and (b) 20 nm iron oxide nanocubes extrapolated from experimental thermal response curves at different maximum applied magnetic fields (Figure S7 within the Supplementary Information). Data are expressed as the mean of 3 measurements ± the standard error of the mean. Inset shows suspension stability after measurement.

**Figure 5. Comparison of Experimental and Computational Results**. SAR values for two nanoparticle solutions of similar concentration (0.5 mg/mL) and size volume but different shape indicating enhancement of SAR values for the 20 nm square nanoparticles. (a) Experimental results. (b) MC simulations for the macrospin model with dipolar interactions at 300K.

**Figure 6. Magnetization simulation.** Hysteresis loops for a spherical (red circles, diameter 20 nm) and a cubic particle (blue squares, side 20 nm) obtained from MC simulations of an atomistic spin model of maghemite at low temperature. In both, uniaxial anisotropy at the core and Néel surface anisotropy have been considered. Snapshots show the spin configurations in the remanent state. Spins have been colored according to their projection into the magnetic field direction (z axis) from red (+1) to blue (-1).

**Figure 7. Chain-like assemblies.** Computed hysteresis loops for arrays of nanoparticles of different length (values of N indicated), and the limit case of a single particle, where $H_A = 2 K/M_S$ is the anisotropy field of the particles. The temperature was introduced in terms of the anisotropy energy barrier of the particles, $t = k_B T/2KV = 0.001$. The inset shows the magnetic response on increasing the temperature for the N = 1 and 10 cases, illustrating the higher thermal stability of the chains. The schematic picture shows particles within a chain possessing easy axes contained within an angle of $\pi/4$ (cubes are only for illustration purposes, since spherical nanoparticles could also form chains).

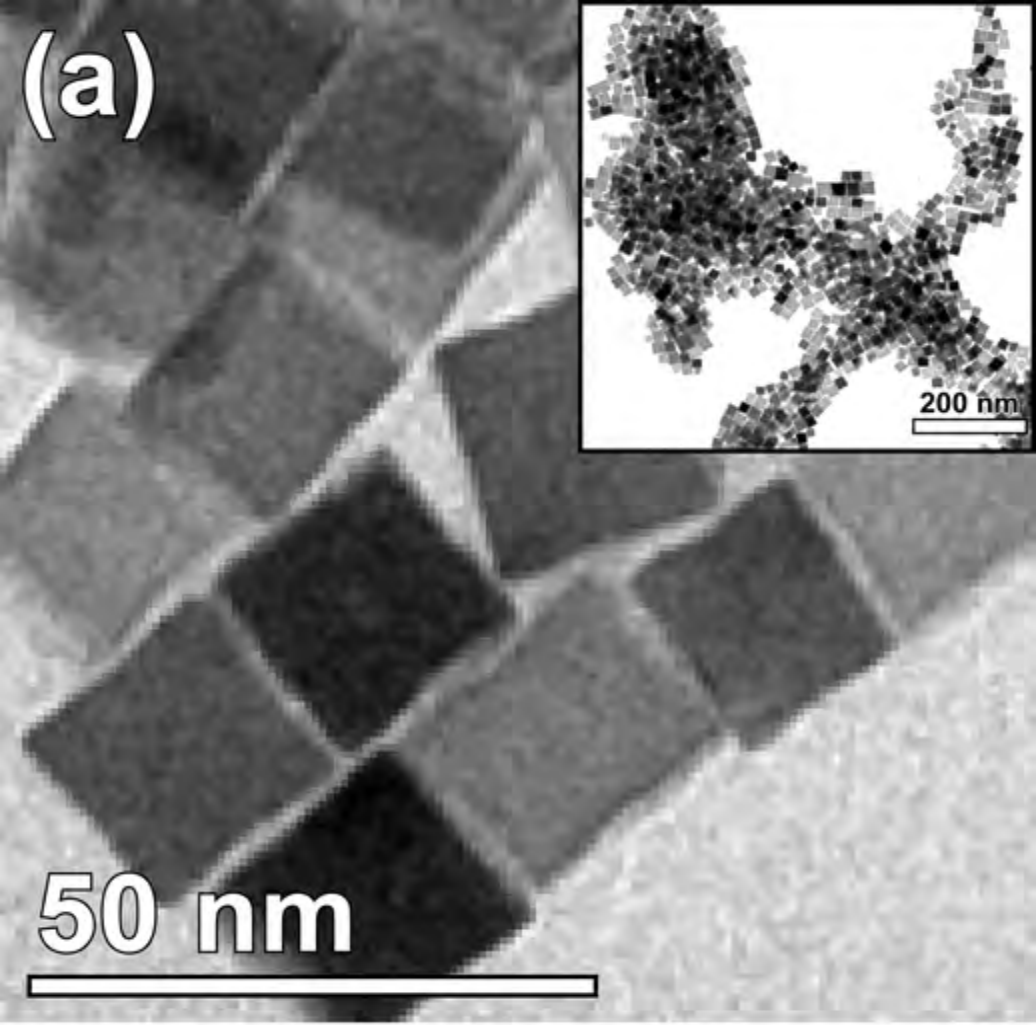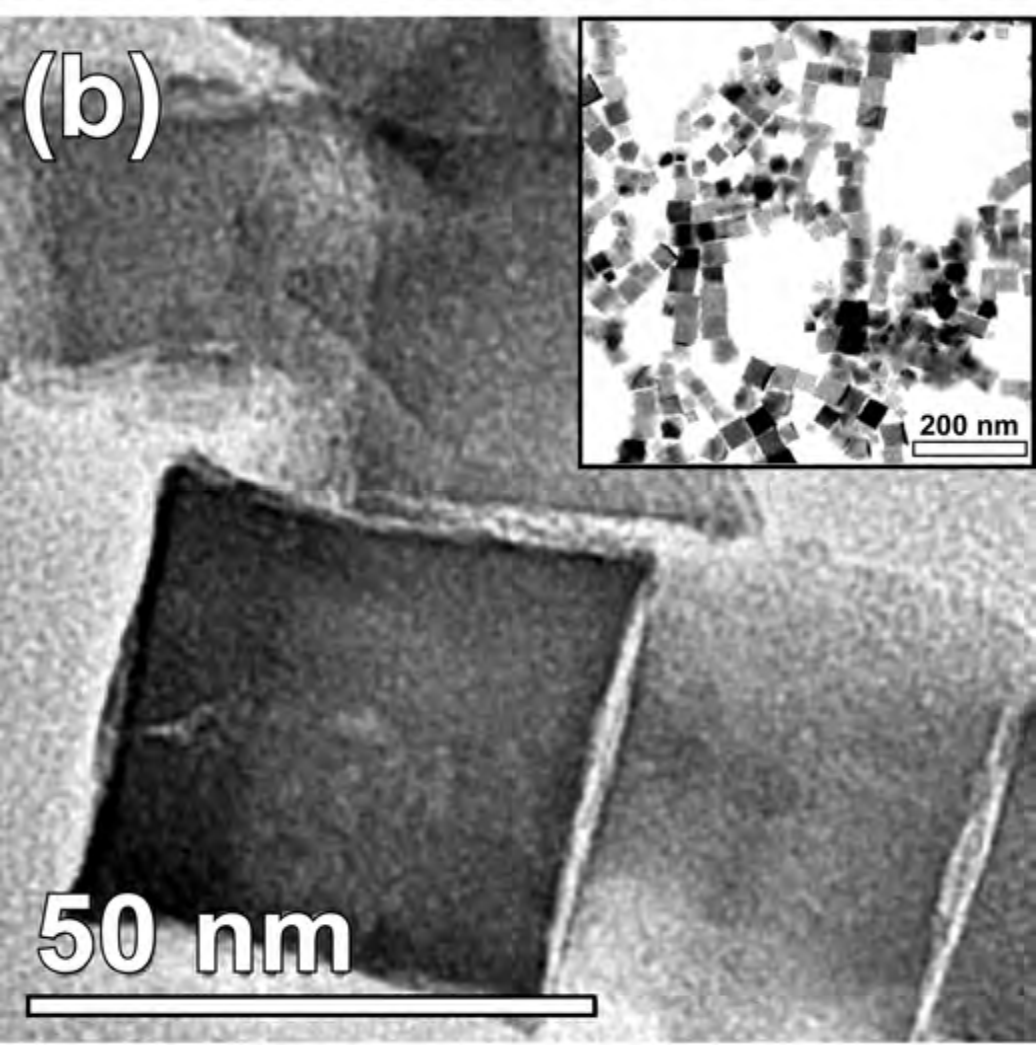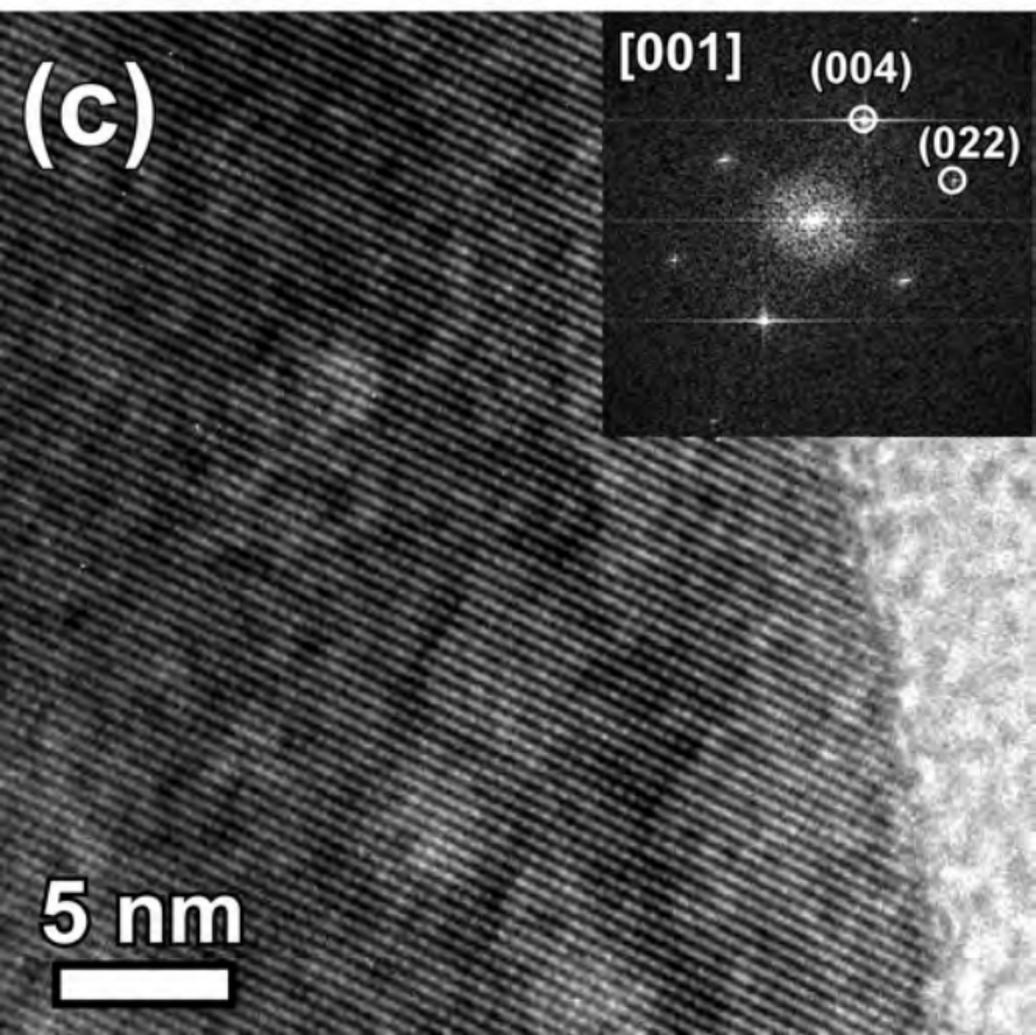

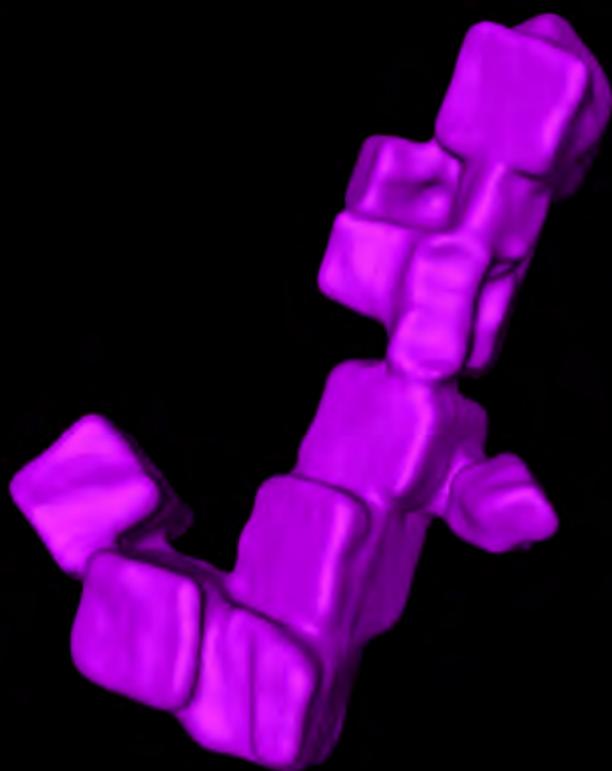

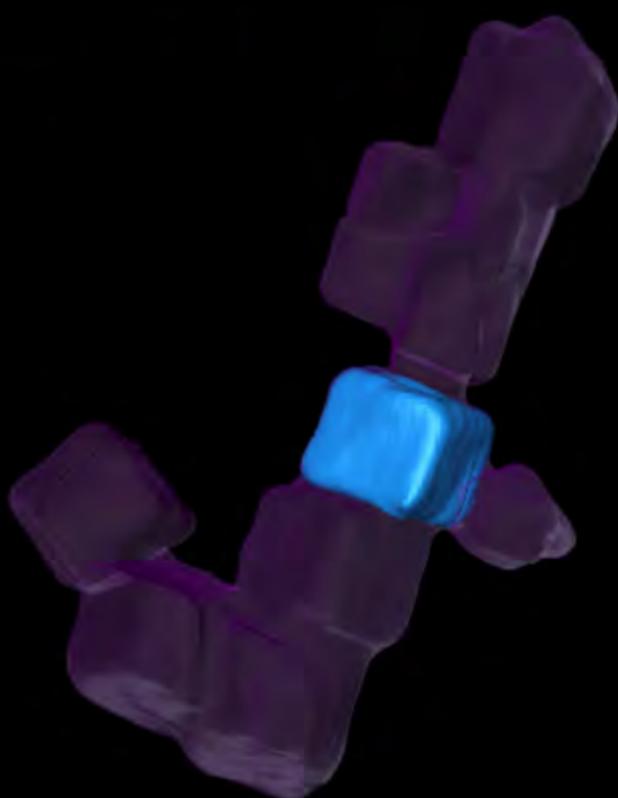

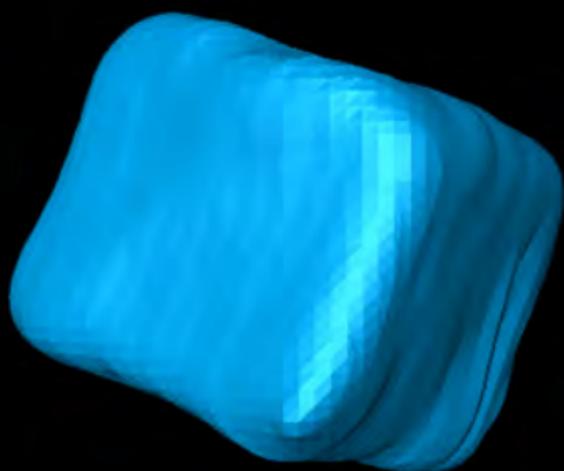

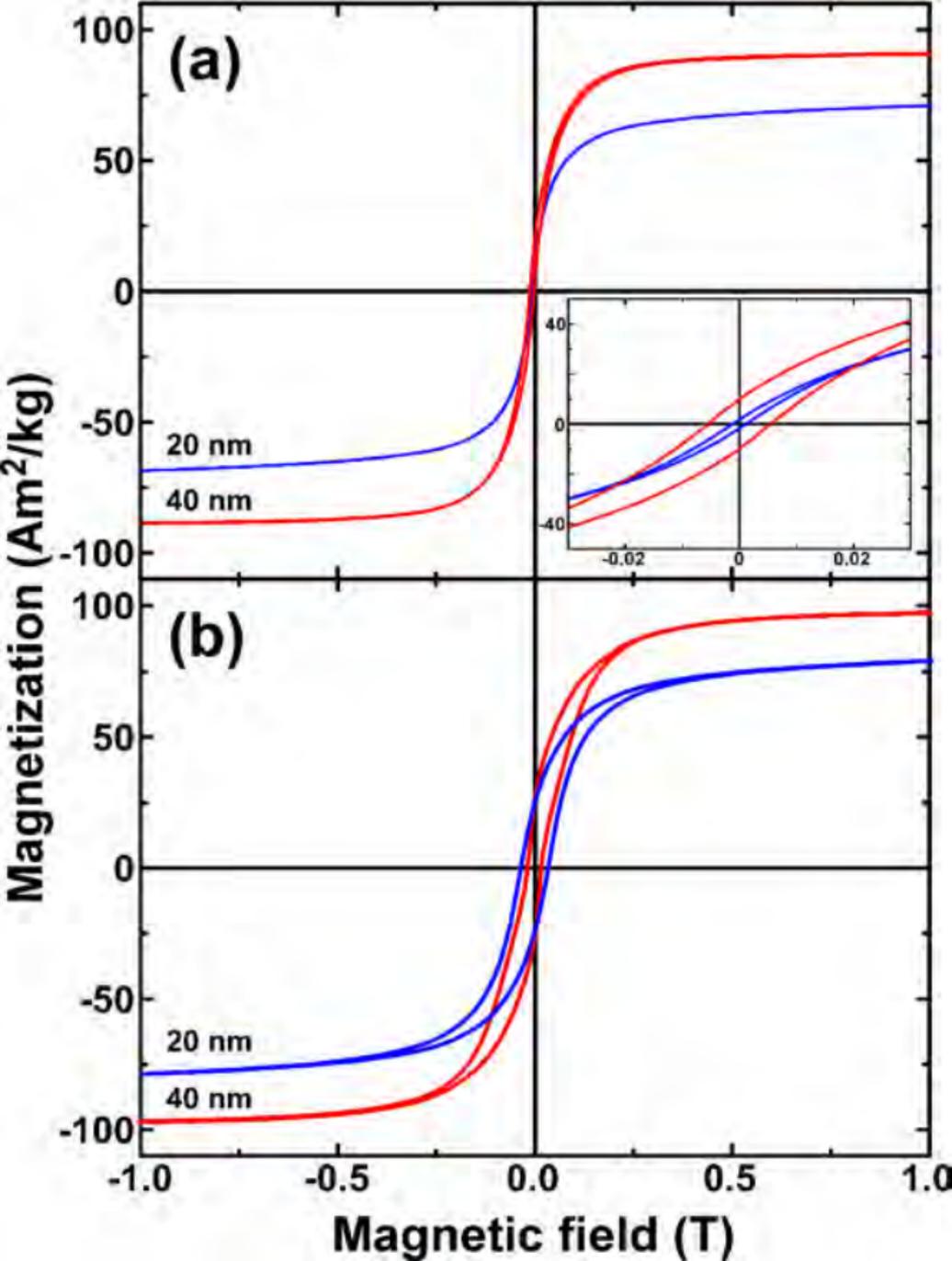

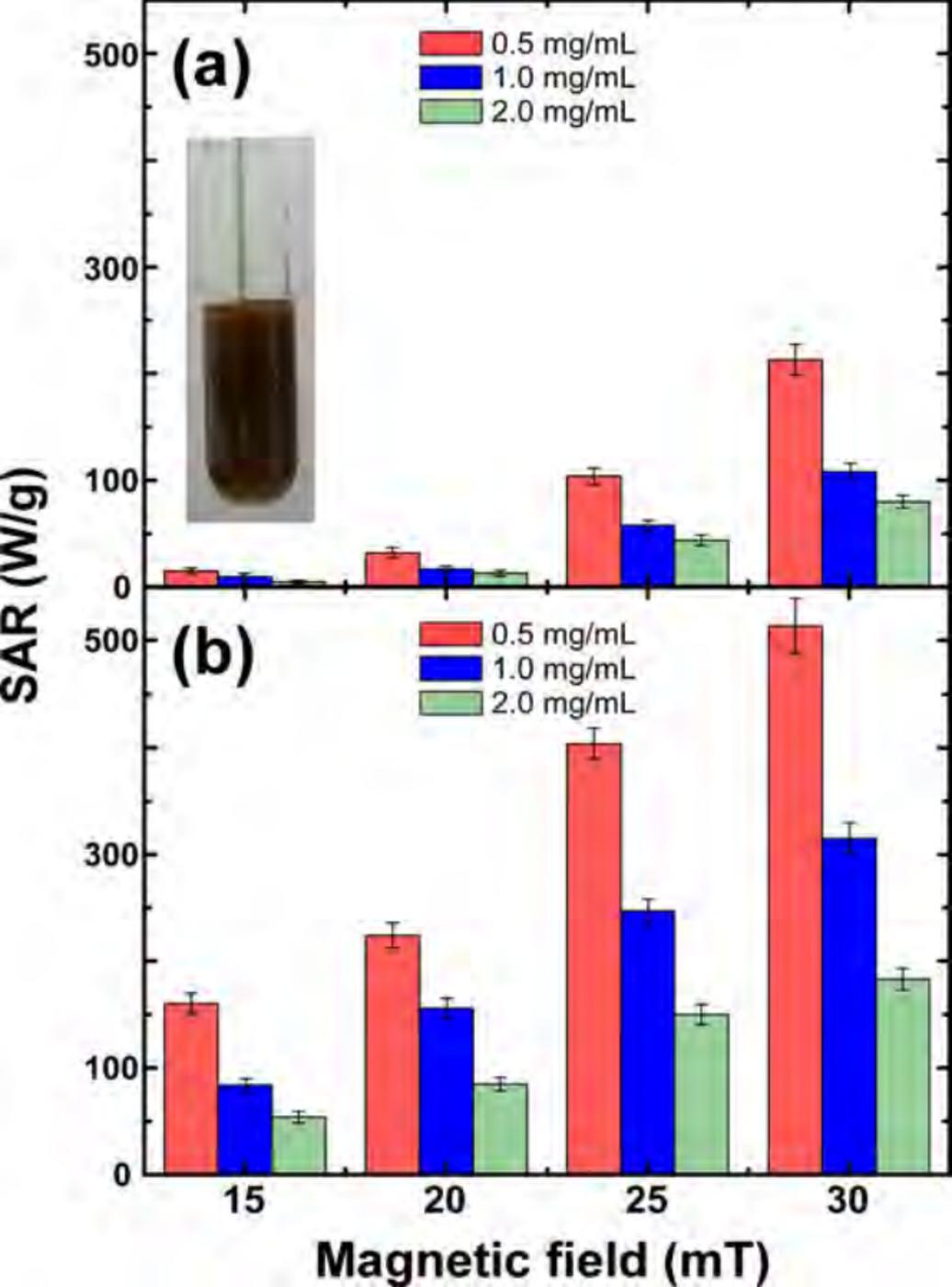

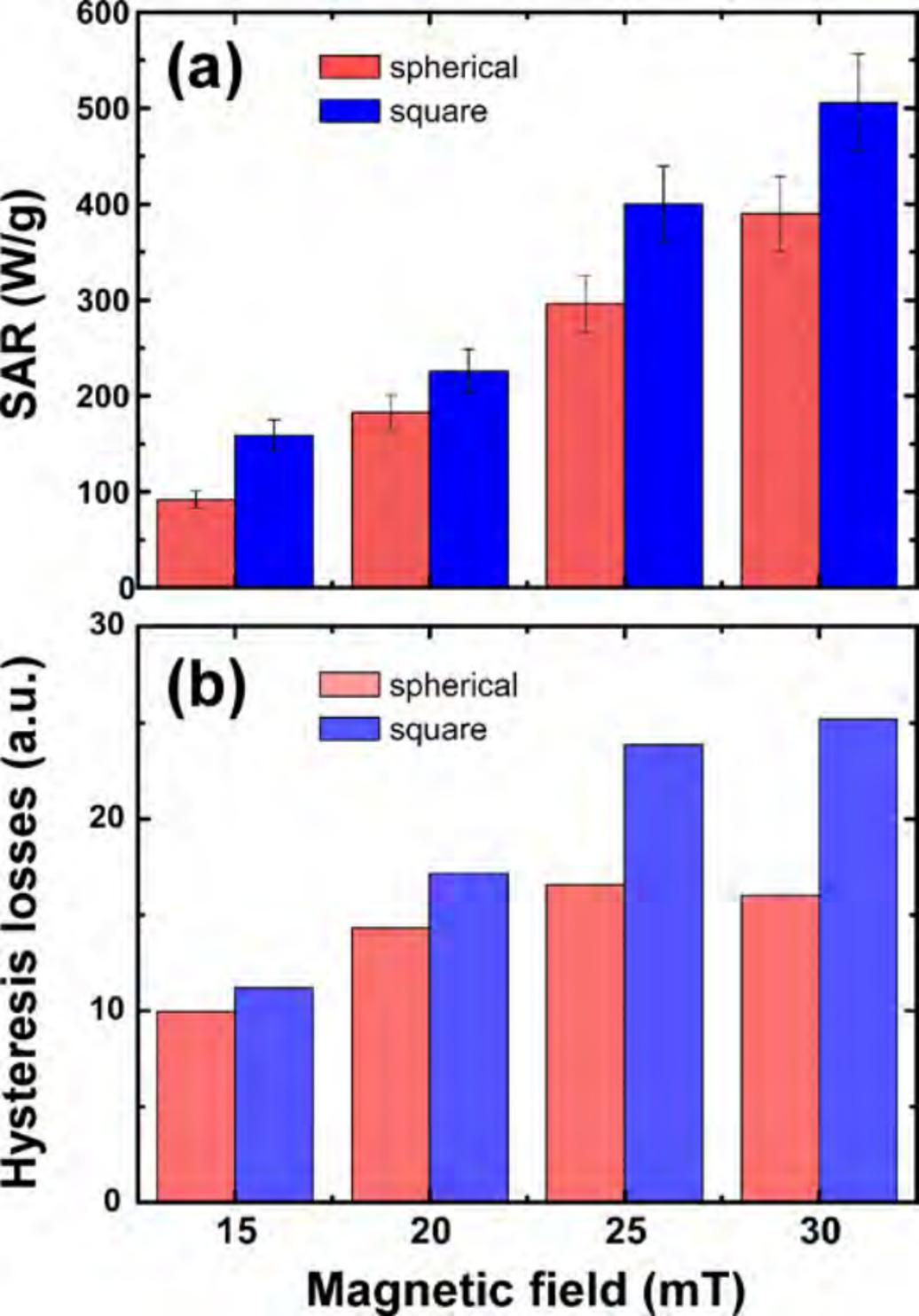

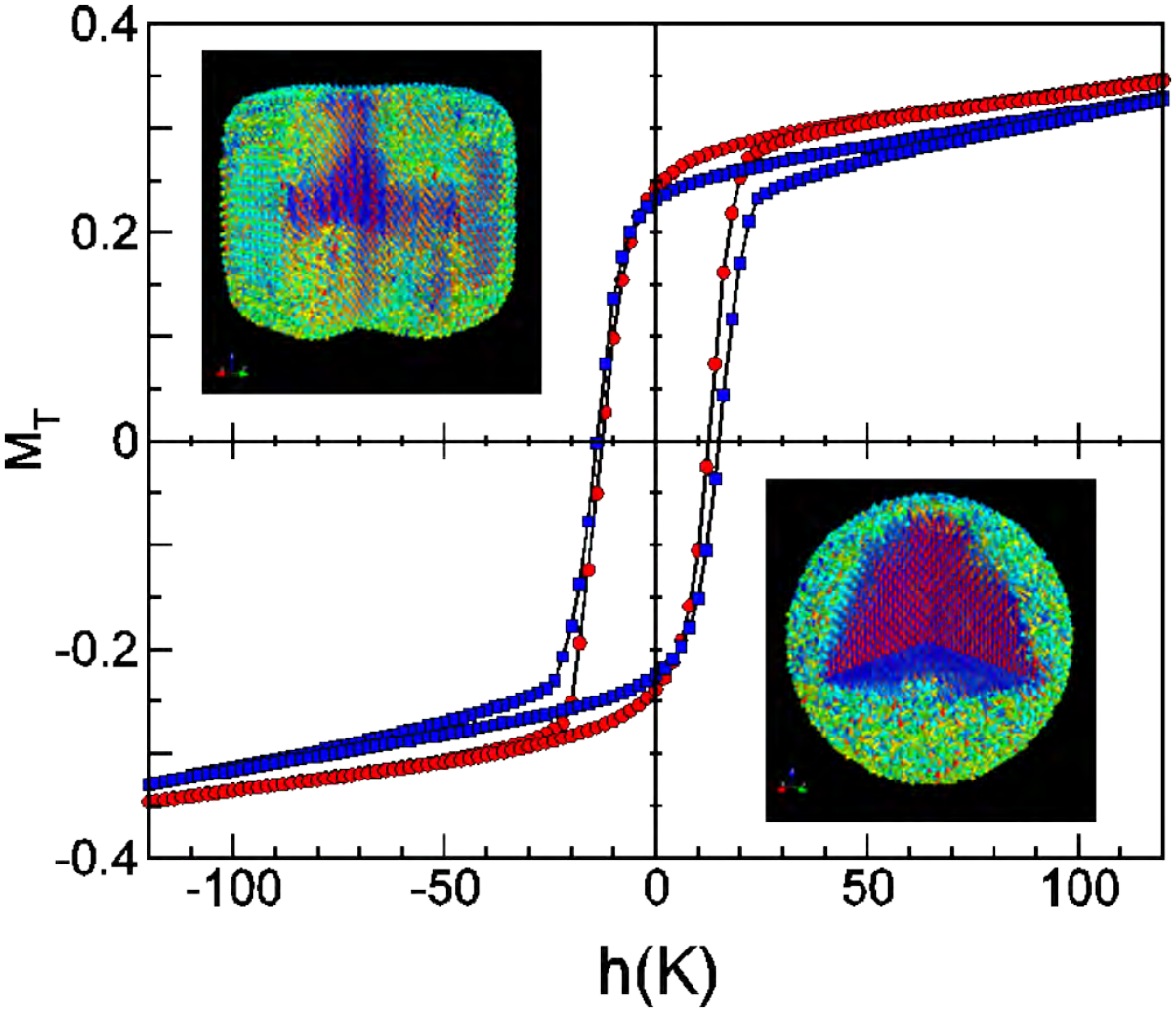

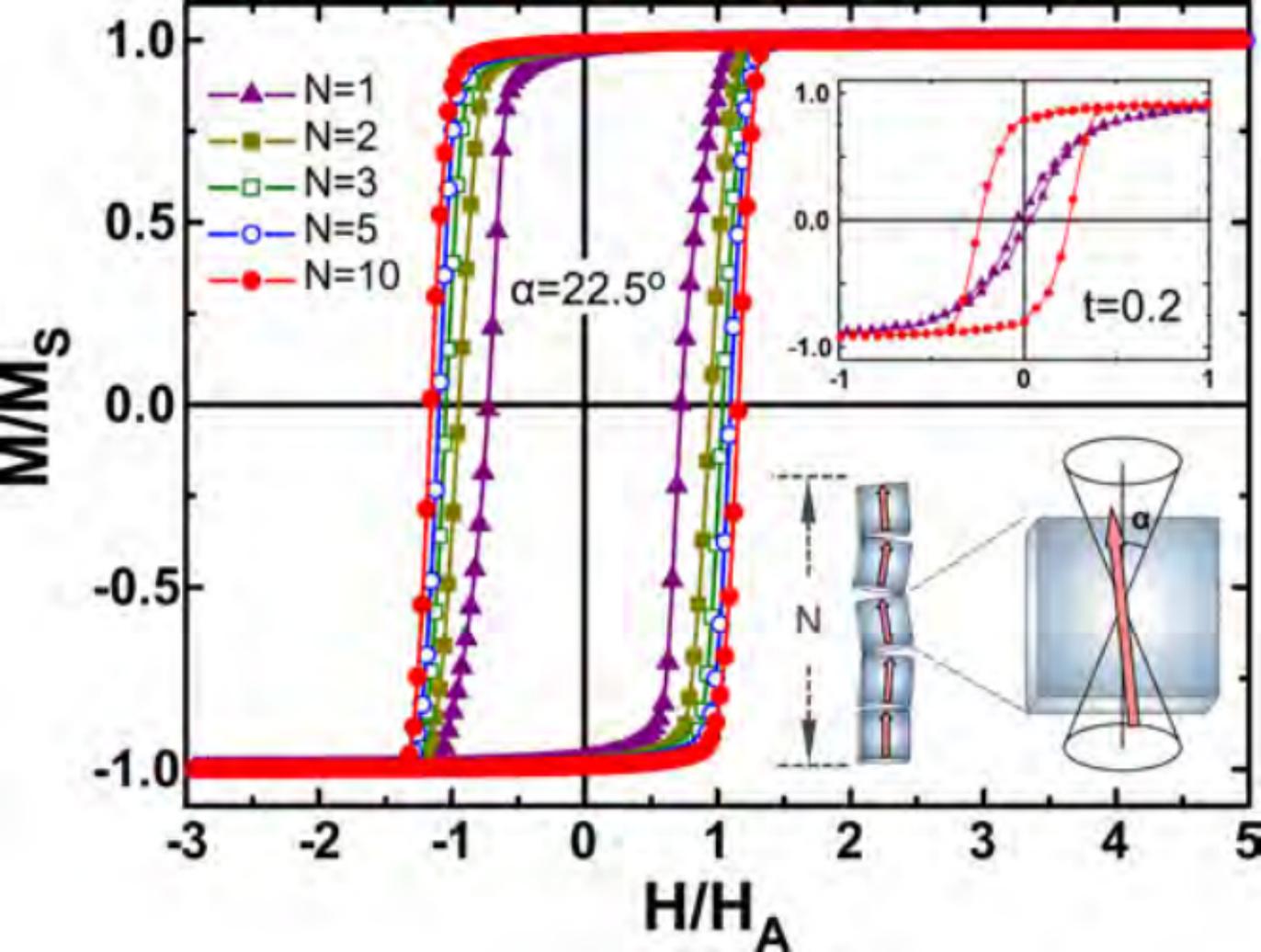

# Learning from Nature to Improve the Heat Generation of Iron-Oxide Nanoparticles for Magnetic Hyperthermia Applications

*Carlos Martinez-Boubeta[1,*], Konstantinos Simeonidis[2], Antonios Makridis[3], Makis Angelakeris[3], Oscar Iglesias[4], Pablo Guardia[5], Andreu Cabot[1,5], Lluis Yedra[1], Sonia Estradé[1,6], Francesca Peiró[1], Zineb Saghi[7], Paul A. Midgley[7], Iván Conde-Leborán[8], David Serantes[8] & Daniel Baldomir[8]*

[1]*Departament d'Electrònica and IN2UB, Universitat de Barcelona, Martí i Franquès 1, 08028 Barcelona, Spain.*

[2]*Department of Mechanical Engineering, University of Thessaly, 38334 Volos, Greece*

[3]*Department of Physics, Aristotle University of Thessaloniki, 54124 Thessaloniki, Greece*

[4]*Departament de Física Fonamental and Institute IN2UB, Universitat de Barcelona, Av. Diagonal 647, 08028 Barcelona, Spain*

[5]*IREC, Jardins de les Dones de Negre 1, 08930 Sant Adrià del Besòs, Spain.*

[6]*TEM-MAT, CCiT-Universitat de Barcelona, Solé i Sabaris 1, 08028 Barcelona, Spain.*

[7]*Department of Materials Science and Metallurgy, University of Cambridge, Pembroke Street, Cambridge CB2 3QZ, United Kingdom.*

[8]*Instituto de Investigacións Tecnolóxicas, and Departamento de Física Aplicada, Universidade de Santiago de Compostela, 15782 Santiago de Compostela, Spain.*

# ADDITIONAL INFORMATION

*MICROSCOPY CHARACTERIZATION*

**Figure S1.** TEM image of the 20 nm spherical nanoparticles.

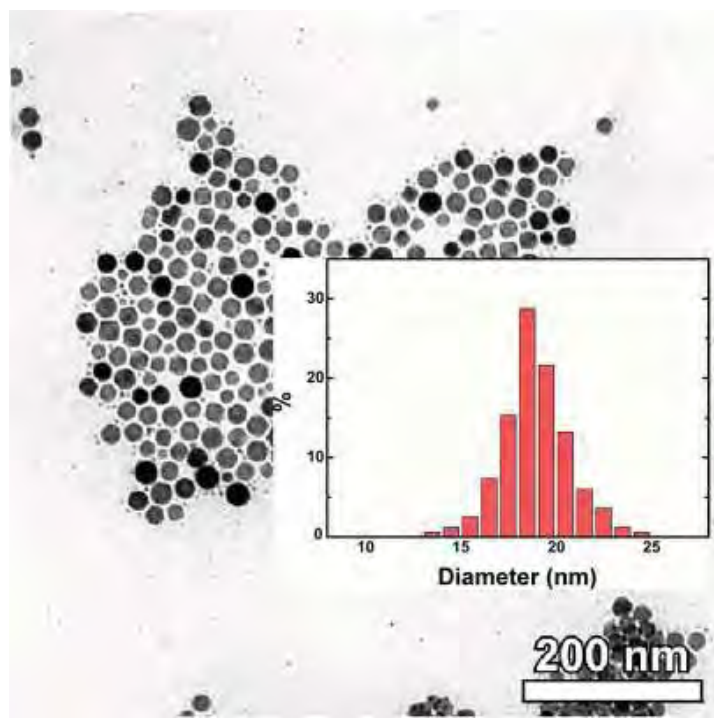

*TOMOGRAPHIC RECONSTRUCTION*

Tomographic reconstruction is achieved using a tilt series covering a range of -64º ≤ α ≤ 64º at an increment of 4º in a FEI Tecnai F20-G2 FEGTEM (200kV). In order to fulfill the projection requirement (Hawkes, P.W. The electron microscope as a structure projector, in: J. Frank (Ed.), Electron Tomography, 2$^{nd}$ ed. Springer-Verlag, **2006,** 83-111), the signal acquired was the high angle annular dark field (HAADF) as the contrast of these images is only proportional to the thickness and the atomic number of the samples. The reconstruction method was the Simultaneous Iterative Reconstruction Technique (SIRT, Gilbert, P. *J. Theor. Biol*. **1972,** *36,* 105-117) with 40 iterations. The reconstructed volume was afterwards manually segmented in order to minimize the elongation effects of the missing wedge of tilt angles.

**Movie S1.** (attached) A movie clip is included in order to visualize the reconstruction and three-dimensional structure.

**Figure S2** presents snapshots of the tomographic recording (on the left) and direct visualization of the reconstruction of the agglomerate volume before segmentation (on the right). From the reconstructed and segmented 3D volume, one particle highlighted in blue was carved out (see Fig. 2 within the main paper). We stress that it is energetically favorable for the (truncated) nanocubes, like in our case, to orient in the face-face configuration due to the {100} facets.

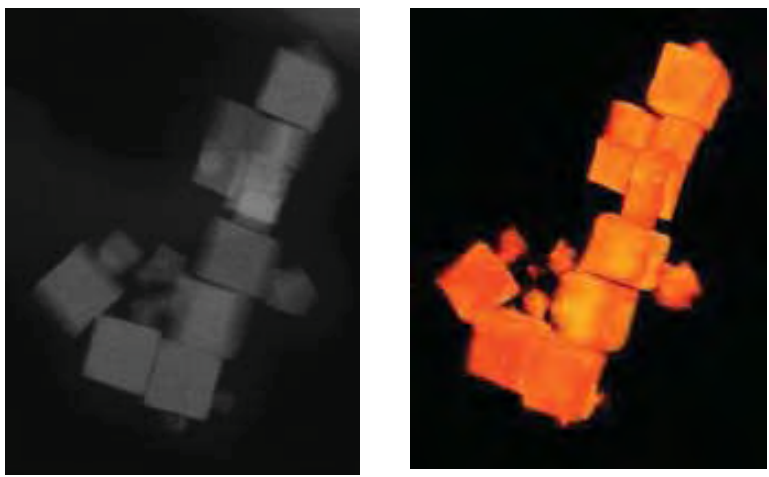

*DYNAMIC LIGHT SCATTERING CHARACTERIZATION*

**Figure S3.** Dynamic light scattering experiments of the colloidal solution displaying the hydrodynamic diameter of the aggregates. Results suggest that cuboids self aggregate in aqueous solution, the average number of particles per chain is roughly 7-10. On the other hand, spherical nanoparticles show mean value $34 \pm 9$ nm, slightly larger than the sizes observed in the TEM images.

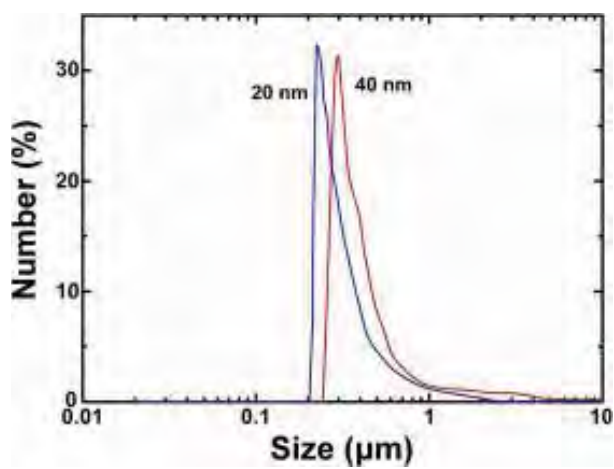

## THERMOGRAVIMETRIC ANALYSIS

**Figure S4.** TGA patterns. A first weight loss of 1-2 %wt. corresponding to moisture occurs below 200 ºC while the gradual decomposition of the bound surfactant molecules takes place between 300-500 ºC. In this range weight decreases about 5 %wt. for 20 nm cuboids and 3 %wt. for 40 nm ones. The theoretical calculation results a percentage of 4.7 and 2.7 respectively for the two samples. The spherical nanoparticles appear to have excess surfactants. On further heating (500-700 ºC range), a lower weight loss is also observed, which could be possibly due to the removal of oxygen atoms from the ferrite lattice by residual carbon [Ayyappan, S. *et al. Mater. Chem. Phys.* **2011**, *130,* 1300].

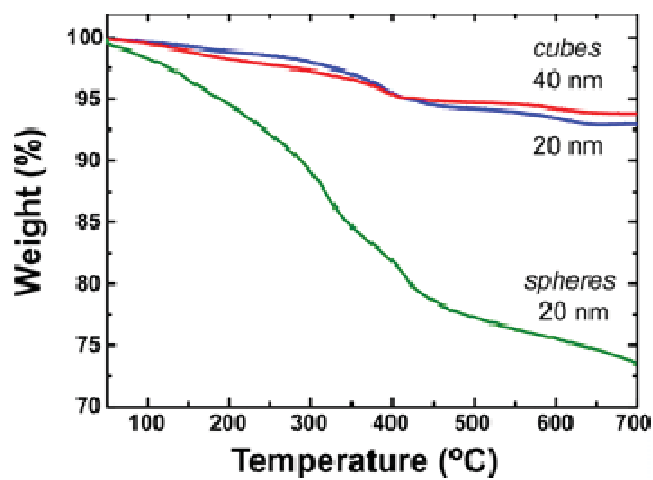

*MAGNETIC CHARACTERIZATION*

**Figure S5.** Hysteresis loops of 20 nm spherical nanoparticles.

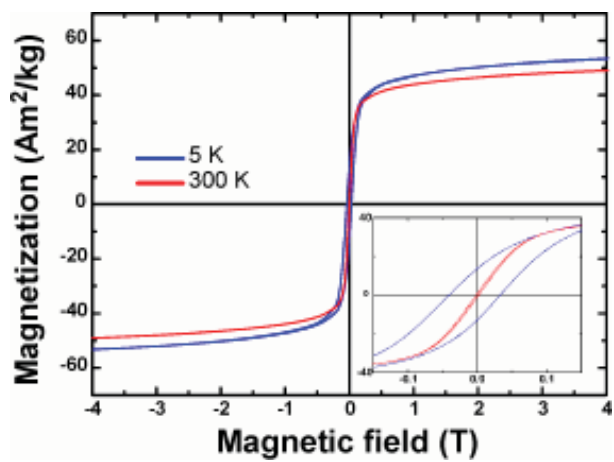

**Figure S6.** Temperature dependence of the magnetization of the iron oxide samples under ZFC and FC conditions. Our previous measurements showed the nanoparticles structure is based on $Fe_3O_4$, but with cation vacancies on decreasing the particle size [Martínez-Boubeta, C. *et al. Phys. Rev. B* **2006**, *74*, 054430]. Accordingly, the 40 nm cubes show incontrovertible evidence for magnetite: the sharp drop in ZFC and FC magnetization near the Verwey transition temperature ($T_V \sim 120$ K), denoted by arrow. Noteworthy, the ZFC curve exhibits a pronounced increase upon heating. Though, above $T_V$ the ZFC and FC curves for the 40 nm sample are similar. Differences arise from the fact the magnetocrystalline anisotropy constant ($K_1$) goes through zero. Thus, from now on, we surmise the 40 nm cuboids are mainly composed of $Fe_3O_4$ while the 20 nm sample resembles maghemite.

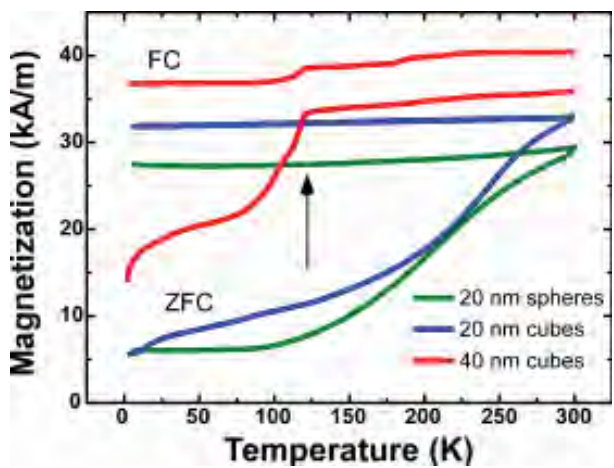

*HYPERTHERMIA MEASUREMENTS*

**Figure S7.** Experimental heating curves, from room temperature, at different iron oxide concentrations. (a-c) 20 nm nanocubes with concentrations ranging from 0.5 mg down to 2 mg of iron oxide per ml, at a frequency of 765 kHz and several magnetic field amplitudes: 15 mT (red line), 20 mT (blue line), 25 mT (black line) and 30 mT (green line). (d-f) Temperature evolution as a function of time for 40 nm cuboids measured under the same experimental conditions as those used in (a-c).

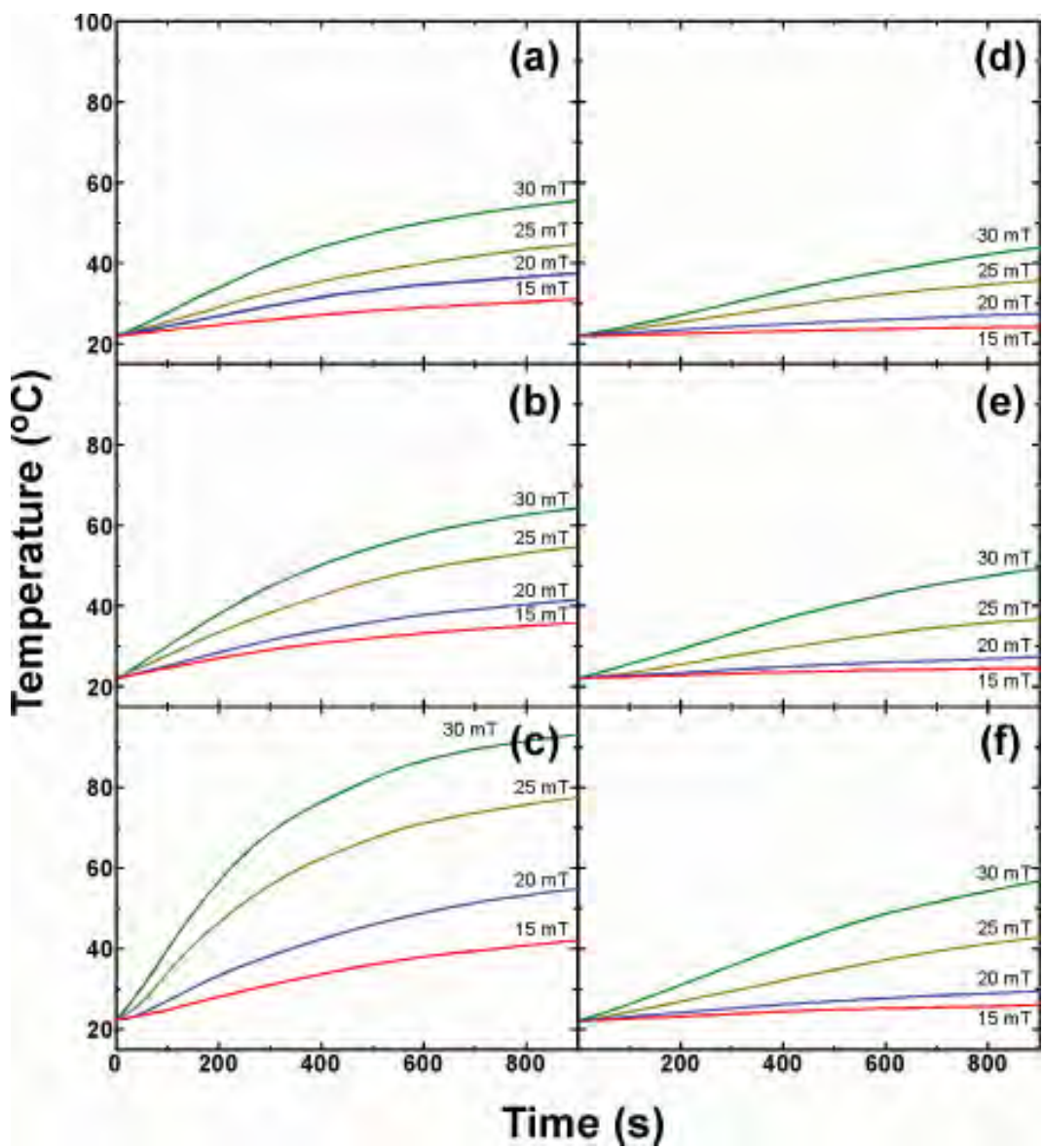

For a safe application of hyperthermia to patients, it was suggested that the product of the frequency and the magnetic field amplitude should be smaller than $5 \times 10^8$ A/ms (Hergt, R.; Dutz, S. *J. Magn. Magn. Mater.* **2007,** *311,* 187). Additionally, a nice survey of literature from January 1990 to October 2010, regarding glioma in vitro studies and assays using animals, can be found in Silva *et al. Int. J. Nanomedicine.* **2011,** *6,* 591. It is observed that the oscillation frequency of the magnetic field, collected for all studies, is mainly defined as being close to 100 kHz. This is also the frequency in use for the first human trials by the company MagForce AG (see for instance Maier-Hauff, *et al. J. Neurooncol.* **2011,** *103,* 317). Yet noteworthy are the first clinical trials ever reported by Gilchrist *et al.* (*Annals of Surgery.* **1957,** *146,* 596), where a frequency of 1.2 MHz and field intensities up to about 300 Oe were used.

While it is admitted that a time-varying magnetic field can cause unpleasant peripheral nerve stimulation when the maximum excursion of the magnetic field is above a frequency-dependent threshold level, please note the aforementioned $H_{max}f$ factor was derived from experiments where 'test person had a sensation of warmth, but was able to withstand the treatment for more than one hour without major discomfort' at low frequency hyperthermia. Obviously, in dependence on the seriousness of the illness this critical $H_{max}f$ product may be exceeded in clinic. For instance, since the eddy current power density is proportional to the square of the diameter of the induced current loops, this criterion would be accordingly weaker for smaller body regions being under field exposure; a combination of field amplitude of about 10 kA/m and frequency of about 400 kHz was suggested for the treatment of breast cancer (Hilger, *et al. J. Magn. Magn. Mater.* **2005,** *293,* 314). In this regard, a very recent study suggests the possibility of new designs for human and preclinical systems, since the degree of unpleasant sensation with magnetic fields up to 0.4 T is shown to decrease significantly above 100 kHz (Weinberg *et al.* **2012,** *Med Phys. 39,* 2578).

On the other hand, please note most importantly clinic applications up-to-date deal with huge amounts of iron oxide nanopowders (see for instance Maier-Hauff *et al. J. Neurooncol.* **2011,** *103,* 317) and may contain concentrations as much as 4 M Fe. In doing so, an uniform heating of the tumour can be obtained even by using low intensity, low frequency fields and non-optimized nanoparticles (see figure below). Though, the possible toxicity and side-effects of this approach need to be explored fully.

**Figure S8** shows that it is possible to obtain the desired temperature increase for hyperthermia therapy even by using non-optimized particles, if present in sufficiently high concentrations. Though, please note the calculated SAR is very low (1 W/g). Again, this is an effect of dipolar interactions that need to be considered when designing real clinic protocols.

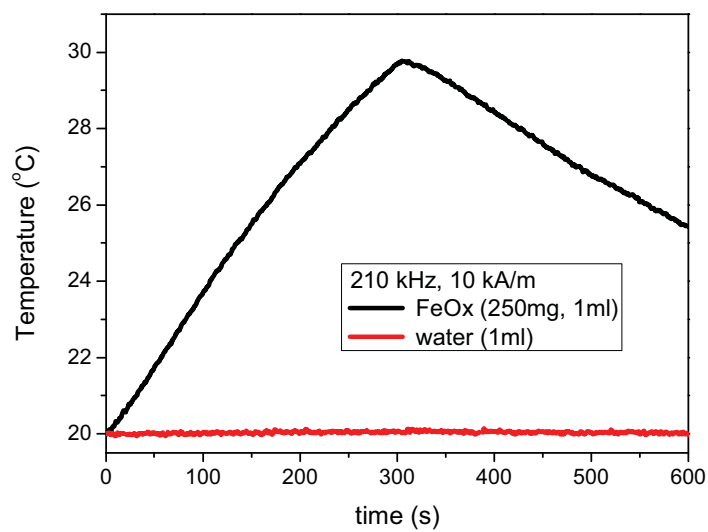

*DETERMINATION OF THE EFFECTIVE MAGNETIC ANISOTROPY CONSTANT*

The temperature dependences of magnetic parameters, such as the coercive field, are interpreted in terms of the changes in the magnetic anisotropy. The reader should please note the values of the magnetocrystalline anisotropy reported in the literature differ considerably; see for instance Abbasi *et al. J. Phys. Chem. C* **2011,** *115,* 6257. Additionally, the effective anisotropy can be larger than the bulk magnetocrystalline anisotropy. For instance, Lartigue *et al. J. Am. Chem. Soc.* **2011***, 133,* 10459 have found decreasing values of $K_{eff}$ on increasing size.

We estimated the effective magnetic anisotropy in the iron oxide particles from the experimental data using the law of approach to magnetic saturation. The high field magnetization curves can be analyzed, at sufficiently high fields, in terms of the law of approach to saturation (see, for instance, Brown, W. F. Jr. Theory of the approach to magnetic saturation. *Phys. Rev.* **1940,** *58,* 736)

$$M = M_S(1 - b/H^2) \qquad (1)$$

where $M_S$ is the saturation magnetization and b is correlated with the effect of the magnetocrystalline anisotropy. Thus, by knowing the fitting parameter b the anisotropy constant can be calculated as $K_{eff} = \mu_0 M_S (15b/4)^{1/2}$ in the case of uniaxial magnetic crystals. The fitting parameters are given below (Table S1). Discrepancies, as it turns out, can be attributed to dipolar coupling between particles (please see also Table S2)

**Table S1.**

| Fitting parameters | | 20 nm cube | 40nm cube | 20nm sphere |
|---|---|---|---|---|
| 5K | $M_S$ [kA m$^{-1}$] | 440 | 505 | 370 |
| | $K_{eff}$ [kJ m$^{-3}$] | 98 | 54 | 15 |
| 300K | $M_S$ [kA m$^{-1}$] | 385 | 465 | 340 |
| | $K_{eff}$ [kJ m$^{-3}$] | 77 | 42 | 12 |

Importantly, we surmise cubic particles exhibit higher anisotropy energy values than that for spheres due to surface contribution. For the cubic nanoparticles, values are higher than those for bulk $Fe_3O_4$ (about 11 kJ m$^{-3}$) or $\gamma$-$Fe_2O_3$ (13 kJ m$^{-3}$) at room temperature and were found to increase with decreasing nanoparticle size. Noteworthy, a similar trend was recently observed by Zhen *et al. J. Phys. Chem. C* **2011,** *115,* 327, but quite the reverse of what Salazar-Alvarez *et al. J. Am. Chem. Soc.* **2008,** *130*, 13234 reported. Those different results could be a consequence of differences in the chemical composition of the nanoparticles with dissimilar sizes and morphology, as already pointed by Chalasani *et al. J. Phys. Chem. C* **2011,** *115,* 18088. In this regard, the presence of core/shell structures within a single particle modifies $K_{eff}$ (see for instance, Shavel *et al. Adv. Func. Mater.* **2007,** *17,* 3870). Nonetheless, in our case since the nanoparticles do not show exchange-bias effects (already reported in Guardia *et al. J. Phys. Chem. C* **2011,** *115,* 390) a homogeneous oxide composition is anticipated. This was further supported by high-resolution TEM (see Fig. 1).

*SIMULATION DETAILS*

We present numerical Monte Carlo calculations using the Metropolis algorithm to simulate the magnetization of interacting ferromagnetic nanoparticles. It is assumed that the nanoparticles are monodomain, with the magnetization change taking place by coherent rotation of the atomic magnetic moments, which means the spin canting effect at the particle surface is not considered. The anisotropy of the particles is of uniaxial type originated from the crystalline contribution, which dominates over the cubic symmetry. First, easy axes directions were distributed at random, as discussed in detail elsewhere (Serantes *et al. J. Appl. Phys.* **2010,** 108, 073918). For each simulation all particles characteristics (magnetic moment, anisotropy, size), are equal, so that we eliminate polydispersity effects in our results. Also, their positions are kept fixed, thus resembling a frozen ferrofluid (reasonable approximation for the particles even inside the cells during hyperthermia treatment). Regarding the computational procedure, the hysteresis losses were determined in the quasistatic limit. Before we proceed we call attention to the fact the frequency and damping parameters have been shown to produce only small alterations to simulation results [Haase, C. & Nowak, U. Role of dipole-dipole interactions for hyperthermia heating of magnetic nanoparticle ensembles. *Phys. Rev. B* **2012,** *85,* 045435].

We used a large system of N = 1000 particles with periodic boundary conditions. In order to estimate the heating power of the system, we simulated M(H) curves with a 10 Oe field variation every 1000 Monte Carlo steps. This field-variation ratio was chosen so that Stoner-Wohlfarth characteristics (coercive field $H_C \approx 0.5H_A$ and remanence $M_R \approx 0.5M_S$) are approximately reproduced at very low temperatures for all particle types, under the conditions of non-interaction and maximum applied field larger than the anisotropy field of the particles. The fitting parameters $M_S$ and volume used for the calculations were independently extracted from the experimental data. Iron oxide density was taken as 5200 kg m$^{-3}$. Regarding K, the exact value of the anisotropy constant is not known for most experimental conditions. Another unknown factor is the possibility of a surface anisotropy in the experimental particles and the effect of the interparticle interactions. Thus, to include the effects of concentration, $K_{eff}$ is adjusted so that the calculated hysteresis curves best fit the experimental SQUID data. The parameters considered accounting for the different types of particles are summarized in Table S2.

**Table S2.** $K_{eff}$ values used in calculations.

| Simulation parameters | | 20 nm cube | 40 nm cube | 20 nm sphere |
|---|---|---|---|---|
| 5 K | $K_{eff}$ [kJ m$^{-3}$] | 26.9 | 14.7 | -- |
| 300 K | $K_{eff}$ [kJ m$^{-3}$] | 3.8 | 2.6 | 3.0 |

Three main aspects are to be investigated: first, the influence of interparticle dipolar interactions on the hyperthermia output of the system, which experimentally are observed to diminish the SAR (see Fig. 4 within the manuscript); second, the observed growing dependence of the SAR on the maximum applied field value (see Figures 4 and 5); and third, the higher hyperthermia output of the 20 nm cubic particles in comparison with the spherical particles of the same volume (d ≈ 25 nm). Our results indicate that, i) the absolute values of K are much smaller than the effective ones estimated from the experiment; ii) the cubic particles have a higher anisotropy, as foreseen from the experimental data and predicted by the atomistic MC simulations (see Fig. 5 and Figure S11); iii) magnetic dipolar interactions decrease the SAR for both the cubic and spherical particles; iv), the SAR grows with $H_{max}$, for the different fields considered. It is remarkable that the experimental trends of the features described above are clearly reproduced, supporting the feasibility of our simulations to study hyperthermia properties.

**Figure S9.** Simulated M(H) curves corresponding to Fig. 4 within the main manuscript. Inset: detail of the coercive fields.

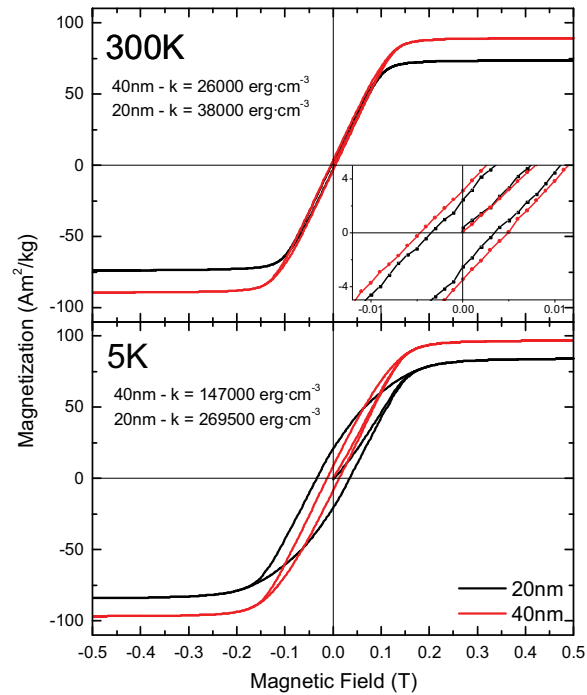

Importantly, data suggest the 20 nm particles at room temperature are in the transition regime between superparamagnetism and a blocked state. In principle this would indicate a lower hyperthermia performance in comparison with particles in the blocked state, for which the hysteresis losses are much more relevant. However, it is worth to note that since the hyperthermia measurements are performed at a much higher frequency ($f \sim 10^5$ Hz), the particle moments will be in that case well below the superparamagnetic transition, ensuring large hysteresis losses. Obviously, the available field-intensities probe only losses from minor hysteresis loops for the 40 nm sample, thus explaining the smaller SAR values compared to the 20 nm case.

At this point it is very important to note that although the simulated values correspond to the energy dissipated in one cycle, it is possible to make a give a good estimation (see Fig. S10) of the results in W/g units, obtaining values very close to the experimental ones. Therefore, to estimate the SAR from the simulations in frequency units, one only has to multiply the hysteresis losses of one cycle times $f$.

**Figure S10.** Calculated area enclosed by hysteresis loops for the 20 nm nanocubes and its concentration dependence.

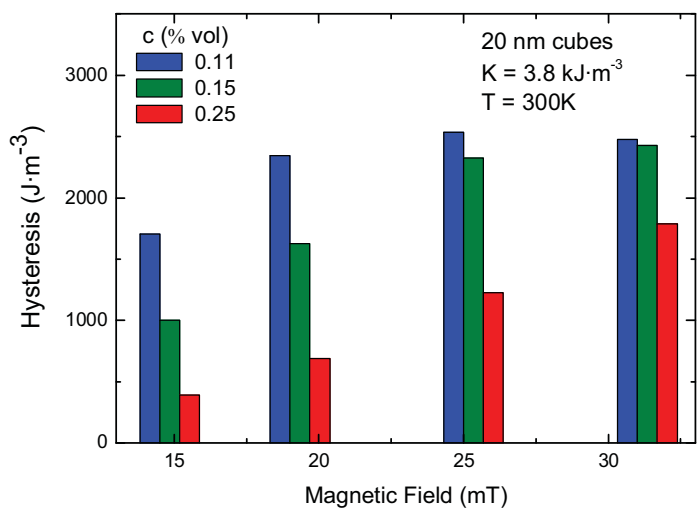

On the other hand, it is well know that a more or less uniformly magnetized state prevails for small spherical nanoparticles. This also remains true for magnetic nanocubes in spite of the divergence of the stray field at the edges and corners (see below). Simulations of individual spherical and cubic nanoparticles were performed on the basis of a system of classical Heisenberg spins by using the Monte Carlo method with the standard Metropolis algorithm. The spins are placed in the positions of the Fe ions of the real maghemite structure as done also in Iglesias, O. & Labarta, A. *Phys. Rev. B* **2001,** *63,* 184416 and then the particles are built by cutting the lattice in a spherical shape or a cube with rounded corners with the same dimensions as in the real samples (12 unit cells diameter/cube length). To compute the energy changes during the simulation, we considered the following Hamiltonian:

$$H/k_B = -\sum_{\langle i,j \rangle} J_{ij}\left(\vec{S}_i \cdot \vec{S}_j\right) - \sum_i \vec{h} \cdot \vec{S}_i + E_{anis} \qquad (2)$$

$$E_{anis} = -k_S \sum_{i \in S} \sum_{j \in nn} \left(\vec{S}_i \cdot \hat{r}_{ij}\right)^2 - k_C \sum_{i \in C} \left(\vec{S}_i \cdot \hat{n}_i\right)^2$$

where the first term accounts for the nn exchange interactions with $J_{ij}$ set to reported values for maghemite, the second is the Zeeman energy with $h = \mu H/k_B$ (H is the magnetic field and μ the magnetic moment of the magnetic ion). The anisotropy energy term $E_{anis}$ distinguishes two kinds of contributions: Néel type anisotropy for the spins with reduced coordination with respect to bulk (surface S spins) and uniaxial anisotropy along the magnetic field direction for the core (C) spins. The standard bulk value is used for $k_C = 4.7 \times 10^5$ J/m$^3$ which translates into a reduced approximate value of $k_C = 0.01$ K/spin. Surface anisotropy constant values were chosen as $k_S = 10$ K for the spherical particle and $k_S = 30$ K for the cubic one, so as to reproduce the effective anisotropies deduced from the experiments. These value were computed by using the well-known relation between anisotropies $k_{eff} = k_C + S/V\, k_S$ (where S and V are the surface and volume of the nanoparticle). Compared with spherical particles we can see both the coercivity and area of the hysteresis loops of cuboids increased (see Fig. S11), so the increased hysteresis losses may give rise to the enhanced heating capacity. For the spherical particle, the smoothness of central area of the loop indicates a quasi-coherent magnetization reversal (as corroborated also by detailed inspection of the snapshots in Fig. S12) with the surface spins progressively aligning towards the field direction as the magnetic field is increased beyond the closure field of the loop. In contrast, the loop of the cubic particle displays a kink-like feature for fields slightly above the coercive fields and presents much higher closure fields. Inspection of the Fig. S11 reveals that the origin of these differential features is in the contribution of surface spins of the lateral faces of the cube that have local anisotropy axis pointing close to the direction perpendicular to the field. Consequently, the peculiar geometry and increased anisotropy of cubic particles translate into an enhanced contribution of surface spins to the coercive and closure fields, producing at the same time non-uniform spin structures during magnetization reversal even below the supposed single domain limit.

**Figure S11.** Contribution of the surface spins only to the hysteresis loop of a spherical (red circles) and a cubic particle with the same characteristics as those in Figure 6.

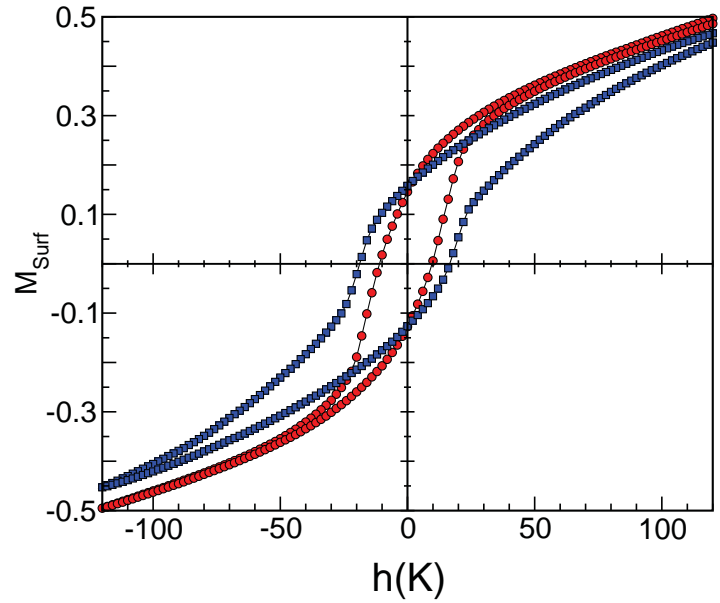

**Figure S12** (on the next page) shows snapshots of the spin configurations at representative points of the hysteresis loops shown in Fig. 6 as indicated: $M_{rem}^{+}$ ($M_{rem}^{-}$), remanent state at the descreasing (increasing) field branch; $H_C^{+}$ ($H_C^{-}$), near the positive (negative) coercive field. Spins have been colored according to their projection into the magnetic field direction (z axis) from red (+1) to blue (-1).

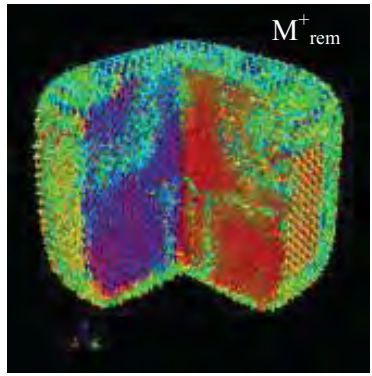
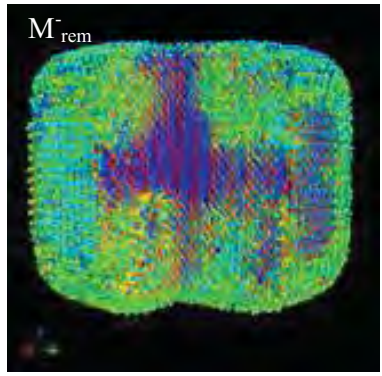
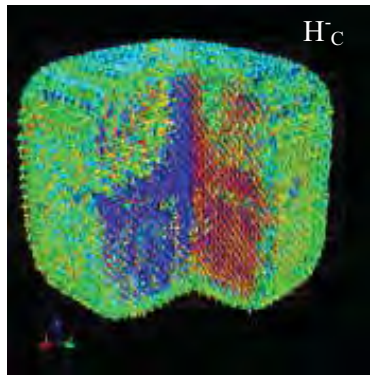
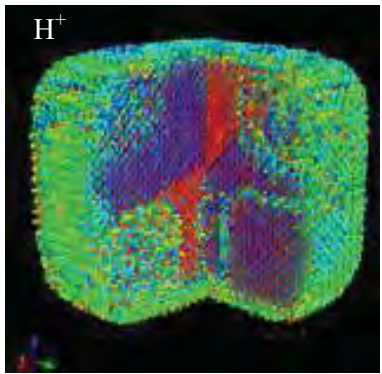
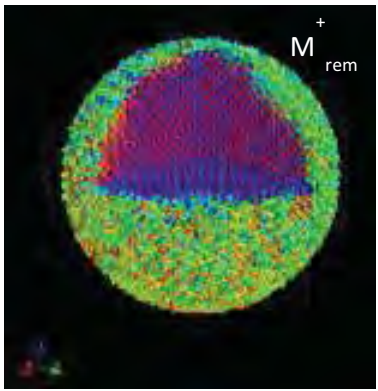
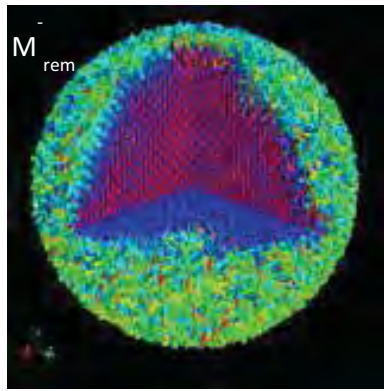
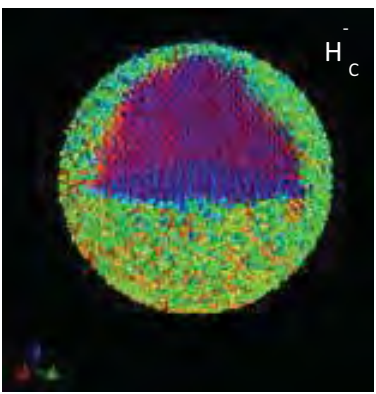
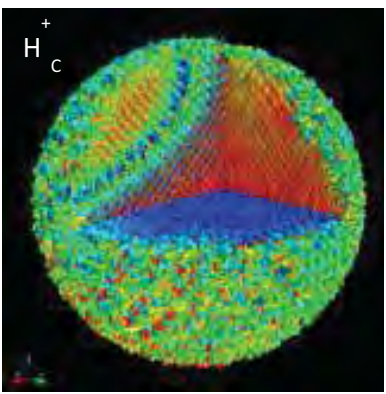